\documentclass[a4paper]{svmult}
\usepackage{times}
\usepackage[dvips]{graphicx,epsfig}
\usepackage{amsmath,amsfonts,amssymb}
\usepackage{cite,url}
\newcommand{\bfdel}{\mbox{\boldmath$\nabla$}}
\begin{document}



\mainmatter

\title*{The Skyrme-Hartree-Fock-Bogoliubov method: its application to finite nuclei and neutron-star crusts}

\author{\underline{{N.~Chamel}}\inst{1}
\and {S.~Goriely}\inst{1}
\and {J.M.~Pearson}\inst{2}}

\titlerunning{The SHFB method: its application to finite nuclei and NS crusts}
\authorrunning{{N.~Chamel}, {S.~Goriely}, and {J.M.~Pearson}}

\toctitle{The Skyrme-Hartree-Fock-Bogoliubov method: its application to finite nuclei and neutron-star crusts}
\tocauthor{{N.~Chamel}, {S.~Goriely},{J.M.~Pearson}}

\institute{{Institut d'Astronomie et d'Astrophysique, Universit\'e
Libre de Bruxelles, CP226,
1050 Brussels, Belgium}
\and {D\'epartement de Physique, Universit\'e de Montr\'eal,
Montr\'eal (Qc) H3C 3J7, Canada}}

\maketitle

\begin{abstract}
After a brief review of the Hartree-Fock-Bogoliubov method with 
Skyrme effective interactions, we show how it can be applied to 
the description of various nuclear systems, from finite nuclei 
to neutron-star crusts. 
\end{abstract}

\section{Introduction}

The global description of properties of finite nuclei over the entire 
nuclear chart requires theoretical methods as microscopic as possible 
and at the same time computationally tractable. It has long been 
recognised that self-consistent mean-field methods with effective 
nucleon-nucleon interactions can be very successfully applied for 
this purpose~\cite{bhr03}. 

In particular, the Brussels-Montreal group has developed a series of 
nuclear mass models based on the Hartree-Fock-Bogoliubov (HFB) method 
with Skyrme effective forces~\cite{pears09}. The model parameters are fitted to 
essentially all the available atomic mass data, with the constraint 
to reproduce several nuclear-matter properties as obtained from microscopic 
calculations using realistic nucleon-nucleon forces. In our latest model HFB-17~\cite{gcp09}, 
we have achieved our best fit ever to essentially all the available 
experimental data, the rms deviation for the set of 2149 measured 
masses of nuclei with $N$ and $Z \ge$ 8~\cite{audi03} being only 
0.581 MeV. Our model was also constrained to fit the 
the equation of state of neutron matter, as calculated by Friedman 
and Pandharipande~\cite{fp81} for realistic two- and three-body 
forces. Besides our model reproduces the $^1S_0$ pairing gaps
in both symmetric nuclear matter and neutron matter from the recent 
Brueckner calculations of Cao et al.~\cite{cao06}. Because of these 
additional constraints, our model can be used to reliably extrapolate 
beyond the neutron drip line and study astrophysical environments 
like for instance the inner crust of neutron stars~\cite{onsi08}.

\section[]{Skyrme-Hartree-Fock-Bogoliubov mass models}

All our HFB mass models are based on a conventional Skyrme force
of the form
\begin{eqnarray}
\label{1}
v^{\rm Sky}(\pmb{r_i}, \pmb{r_j})  &=&  t_0(1+x_0 P_\sigma)\delta({\pmb{r}_{ij}}) \nonumber \\
& &+\frac{1}{2} t_1(1+x_1 P_\sigma)\frac{1}{\hbar^2}\left[p_{ij}^2\,\delta({\pmb{r}_{ij}})
+\delta({\pmb{r}_{ij}})\, p_{ij}^2 \right]\nonumber\\
& &+t_2(1+x_2 P_\sigma)\frac{1}{\hbar^2}\pmb{p}_{ij}.\delta(\pmb{r}_{ij})\,
 \pmb{p}_{ij} \nonumber \\
& &+\frac{1}{6}t_3(1+x_3 P_\sigma)\rho(\pmb{r})^\gamma\,\delta(\pmb{r}_{ij})
\nonumber\\
& &+\frac{\rm i}{\hbar^2}W_0(\mbox{\boldmath$\sigma_i+\sigma_j$})\cdot
\pmb{p}_{ij}\times\delta(\pmb{r}_{ij})\,\pmb{p}_{ij}  \quad ,
\end{eqnarray}
where $\pmb{r}_{ij} = \pmb{r}_i - \pmb{r}_j$, $\pmb{r} = (\pmb{r}_i + 
\pmb{r}_j)/2$, $\pmb{p}_{ij} = - {\rm i}\hbar(\pmb{\nabla}_i-\pmb{\nabla}_j)/2$
is the relative momentum, and $P_\sigma$ is the two-body 
spin-exchange operator. 
Following the usual practice, we consider a different 
force in the pairing channel. The latter acts only between nucleons of the same 
charge state $q$ ($q = n$ or $p$ for neutron or proton, respectively) and is given by
\begin{equation}
v^{\rm pair}_q(\pmb{r_i}, \pmb{r_j})= v^{\pi\,q}[\rho_n(\pmb{r}),\rho_p(\pmb{r})]~\delta(\pmb{r}_{ij})\, ,
\label{2}
\end{equation}
where $v^{\pi\,q}[\rho_n,\rho_p]$ is a functional of the nucleon densities. 

Assuming time-reversal invariance, the ground-state energy can be written as 
the integral of a purely local energy-density functional 
$\mathcal{E}_{\rm HFB}(\pmb{r})$ which depends on 
\noindent (i) the nucleon density (denoting the spin states by $\sigma=\pm1$),
\begin{equation}
\rho_q(\pmb{r}) = \sum_{\sigma=\pm 1}\rho_q(\pmb{r}, \sigma; \pmb{r}, \sigma)
\, ,
\end{equation}
(ii) the kinetic-energy density (in units of $\hbar^2/2M_q$),
\begin{equation}
\tau_q(\pmb{r}) = \sum_{\sigma=\pm 1}\int\,{\rm d}^3\pmb{r^\prime}\,\delta(\pmb{r}-\pmb{r^\prime}) \bfdel\cdot\bfdel^\prime
\rho_q(\pmb{r}, \sigma; \pmb{r^\prime}, \sigma)
\end{equation}
(iii) the spin-current density,
\begin{eqnarray}
\pmb{J}_q(\pmb{r}) = -{\rm i}\sum_{\sigma,\sigma^\prime=\pm1}\int\,{\rm d}^3\pmb{r^\prime}\,\delta(\pmb{r}-\pmb{r^\prime})
\bfdel\rho_q(\pmb{r}, \sigma; \pmb{r^\prime},
\sigma^\prime) \times \mbox{\boldmath$\sigma$}_{\sigma^\prime \sigma}   \nonumber \\
={\rm i}\sum_{\sigma,\sigma^\prime=\pm1}\int\,{\rm d}^3\pmb{r^\prime}\,\delta(\pmb{r}-\pmb{r^\prime})
\bfdel^\prime\rho_q(\pmb{r}, \sigma; \pmb{r^\prime},
\sigma^\prime) \times \mbox{\boldmath$\sigma$}_{\sigma^\prime \sigma}
\end{eqnarray}
and (iv) the abnormal density,
\begin{equation}
\tilde{\rho}_q(\pmb{r}) = \sum_{\sigma=\pm 1}
\tilde{\rho}_q(\pmb{r}, \sigma ; \pmb{r}, \sigma)   \, ,
\end{equation}
where $\mbox{\boldmath$\sigma$}_{\sigma\sigma^\prime}$ denotes the Pauli spin
matrices. In turn the normal and abnormal density matrices, 
$\rho(\pmb{r}, \sigma; \pmb{r^\prime}, \sigma^\prime)$ and 
$\tilde{\rho}(\pmb{r}, \sigma; \pmb{r^\prime}, \sigma^\prime)$ respectively, 
can be expressed as~\cite{doba84,doba96}
\begin{equation}
\rho_q(\pmb{r}, \sigma; \pmb{r^\prime}, \sigma^\prime) =
\sum_{i(q)}\psi^{(q)}_{2i}(\pmb{r}, \sigma)\psi^{(q)}_{2i}(\pmb{r^\prime}, \sigma^\prime)^* 
\end{equation}
and
\begin{equation}
\tilde{\rho}_q(\pmb{r}, \sigma; \pmb{r^\prime}, \sigma^\prime) =
-\sum_{i(q)}\psi^{(q)}_{2i}(\pmb{r}, \sigma)
\psi^{(q)}_{1i}(\pmb{r^\prime}, \sigma^\prime)^*=-
\sum_{i}\psi^{(q)}_{1i}(\pmb{r}, \sigma)\psi^{(q)}_{2i}(\pmb{r^\prime},
\sigma^\prime)^* \, ,
\end{equation}
where $\psi^{(q)}_{1i}(\pmb{r}, \sigma)$ and $\psi^{(q)}_{2i}(\pmb{r}, \sigma)$ 
are the two components of the quasiparticle wavefunction. 
Minimizing the HFB energy with respect to $\psi^{(q)}_{1i}(\pmb{r}, \sigma)$ and 
$\psi^{(q)}_{2i}(\pmb{r}, \sigma)$ under the constraints of fixed particle numbers 
leads to the HFB equations~\cite{doba84,doba96}
\begin{eqnarray}
\label{5}
\sum_{\sigma^\prime}
\begin{pmatrix} h^\prime_q(\pmb{r} )_{\sigma \sigma^\prime} & \Delta_q(\pmb{r}) \delta_{\sigma \sigma^\prime} \\ \Delta_q(\pmb{r}) \delta_{\sigma \sigma^\prime} & -h^\prime_q(\pmb{r})_{\sigma \sigma^\prime} 
 \end{pmatrix}\begin{pmatrix} 
\psi^{(q)}_{1i}(\pmb{r},\sigma^\prime) \\ \psi^{(q)}_{2i}(\pmb{r},\sigma^\prime) \end{pmatrix} = \nonumber\\
\begin{pmatrix} E_i+\lambda_q & 0 \\ 0 & E_i-\lambda_q \end{pmatrix}
\begin{pmatrix} \psi^{(q)}_{1i}(\pmb{r},\sigma) \\ \psi^{(q)}_{2i}(\pmb{r},\sigma) \end{pmatrix}
\end{eqnarray}
where $\lambda_q$ are Lagrange multipliers. 
The single particle Hamiltonian 
$h^\prime_q(\pmb{r} )_{\sigma \sigma^\prime}$ and pairing field $\Delta_q(\pmb{r})$  are given by 
\begin{equation}
\label{6}
h^\prime_q(\pmb{r})_{\sigma^\prime\sigma} \equiv -\bfdel\cdot
B_q(\pmb{r})\bfdel\, \delta_{\sigma\sigma^\prime}
+ U_q(\pmb{r}) \delta_{\sigma\sigma^\prime}
-{\rm i}\pmb{W_q}(\pmb{r}) \cdot\bfdel\times\mbox{\boldmath$\sigma$}_{\sigma^\prime\sigma}
\end{equation}
and
\begin{equation}
\label{7}
\Delta_q(\pmb{r})=\frac{1}{2}v^{\pi q} [\rho_n(\pmb{r}),\rho_p(\pmb{r})]\tilde{\rho}_q(\pmb{r}) \, .
\end{equation}
The single particle fields appearing in Eq.~(\ref{6}) are defined by
\begin{eqnarray}
B_q(\pmb{r}) &=&
\frac{\partial \mathcal{E}_{\rm HFB}(\pmb{r})}{\partial\tau_q(\pmb{r})}\, ,
\hskip0.5cm
U_q(\pmb{r})=\frac{\partial \mathcal{E}_{\rm HFB}(\pmb{r})}{\partial\rho_q(\pmb{r})} - \pmb{\nabla}\cdot\frac{\partial \mathcal{E}_{\rm HFB}(\pmb{r})}{\partial (\pmb{\nabla}\rho_q(\pmb{r}))}\, ,\nonumber \\
\pmb{W}_q(\pmb{r})&=&\frac{\partial \mathcal{E}_{\rm HFB}(\pmb{r})}
{\partial\pmb{J}_q(\pmb{r})}  \, .
\end{eqnarray}
Expressions for these fields can be found 
for instance in Ref.~\cite{cha08}. In the absence of pairing, the HFB equations~(\ref{5}) reduce to the Hartree-Fock equations. 

In homogeneous matter the HFB equations~(\ref{5}) can be readily solved. In particular, the pairing field is uniform and 
obey the well-known BCS gap equations (see for instance Appendix B of Ref.~\cite{cha08})
\begin{equation}
\label{8}
\Delta_q=- \frac{1}{8\pi^2} B_q^{3/2} v^{\pi\, q}[\rho_n,\rho_p] \, \Delta_q \,\int_\Lambda {\rm d}\varepsilon 
\frac{\sqrt{\varepsilon}}{\sqrt{(\varepsilon-\lambda_q)^2+\Delta_q^2}} \, ,
\end{equation}
where the subscript $\Lambda$ is to indicate that the integral has to be 
regularized by introducing a cutoff. 
In our latest models HFB-16~\cite{cha08} and HFB-17~\cite{gcp09}, we have inverted the gap equation~(\ref{8}) at 
each neutron and proton density in order to deduce the effective pairing strength $v^{\pi\, q}[\rho_n,\rho_p]$ 
from a given gap function $\Delta_q(\rho_n,\rho_p)$. In particular, for the model HFB-17 we have taken 
the $^1S_0$ pairing gap obtained from many-body calculations including medium polarisation effects and 
using realistic two- and three-body forces~\cite{cao06}. 

\begin{figure}[t]\centering
\includegraphics[width=100mm]{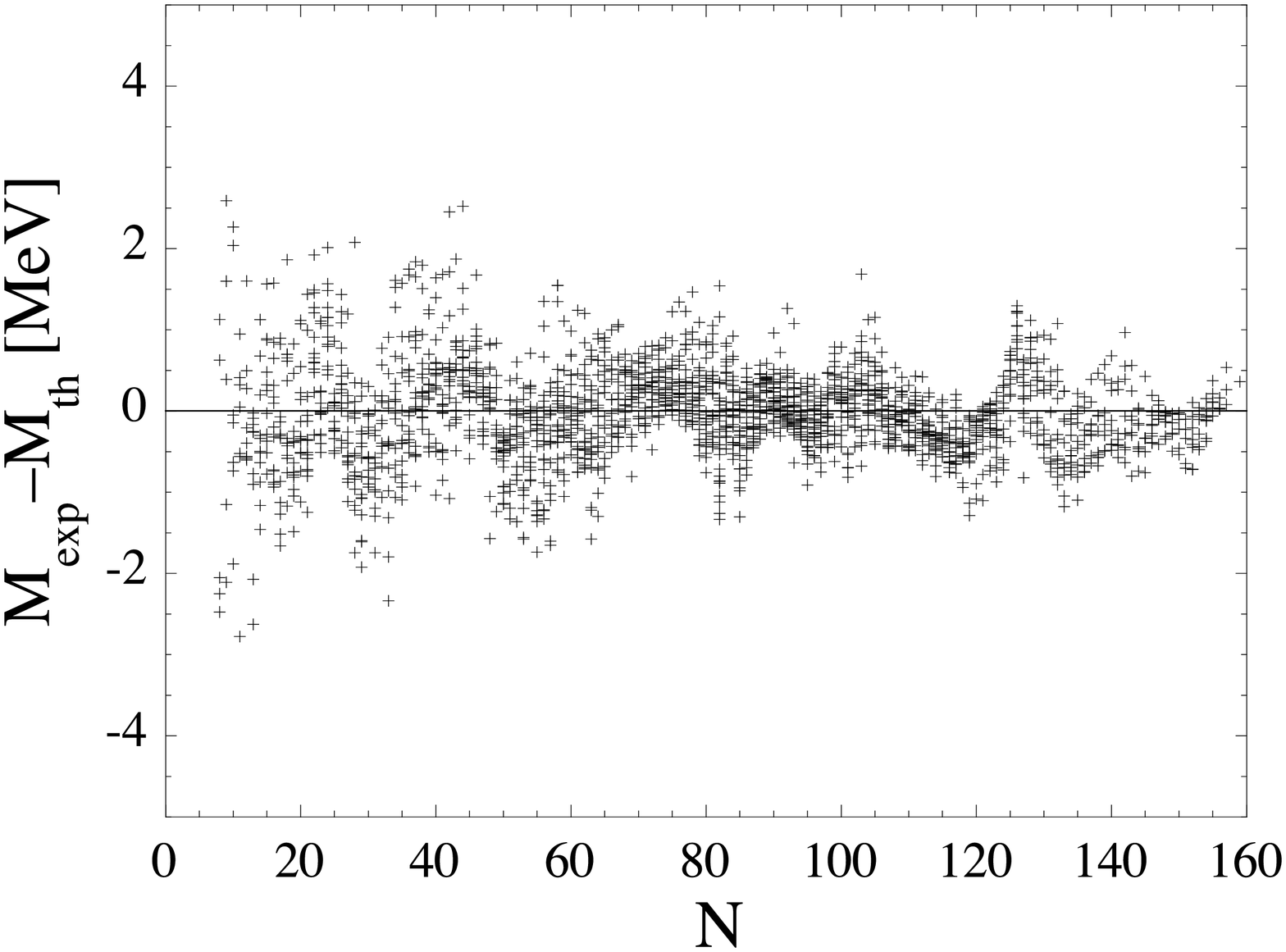}
\caption{Differences between experimental and calculated masses as a function 
of the neutron number $N$ for the HFB-17 mass model.}
\label{fig:1}
\end{figure}

For applications to nuclear masses, two phenomenological corrections are added to the HFB ground-state energy: 
(i) a Wigner energy and (ii) a rotational and vibrational spurious collective energy (see for instance
Ref.~\cite{cha08} for details). The deviations between the 2149 measured masses of nuclei with $N$ and $Z \ge$ 8 given in 
the 2003 AME~\cite{audi03} and the predictions from our model HFB-17 are shown graphically in Fig.~\ref{fig:1}. 
The rms and mean values of these deviations are  0.581~MeV and -0.019~MeV, respectively. 
HFB-17 is the most accurate mass model ever achieved within the mean-field framework. The complete HFB-17 table 
of 8389 masses including all nuclei with $Z$,$N \geq 8$ and $Z\leq 110$ and lying between the proton and the 
neutron drip lines is available on our website\footnote{\url{http://www-astro.ulb.ac.be}}.
With the additional constraints on infinite nuclear matter, our HFB mass model is also particularly 
suitable for astrophysics applications such as the description of neutron-star crusts~\cite{onsi08}. 

\section[]{Applications to neutron stars}

Neutron stars are among the most compact objects in the Universe with a central density 
which can exceeds several times that found inside heavy atomic nuclei. Neutron stars are 
born in the catastrophic gravitational core collapse of massive stars in supernova explosions. 
The outer layers of the star are formed of a solid crust~\cite{lrr}, which at densities below the neutron
 drip threshold $\rho_{\rm ND}\simeq 4\times 10^{11}$ g.cm$^{-3}$, is composed of a solid Coulomb
lattice of neutron-rich nuclei with $Z/A\lesssim 0.5$ coexisting with a degenerate gas of relativistic electrons.
Our HFB mass models can be directly used to compute the composition and the equation of state of these
layers following the classical work of Baym, Pethick and Sutherland~\cite{bps71}. 
We have found essentially the same sequence of nuclides (see for instance Table 4 of Ref.~\cite{pears09}) 
than that obtained by Haensel and Pichon~\cite{haen94}. 
In particular the last equilibrium nuclide at densities just below the neutron drip transition is $^{120}$Sr. 
The inner crust of neutron stars, at densities above $\rho_{\rm ND}$ up to about half the saturation 
density is permeated by a neutron ocean. The latter affects the properties of the ``nuclei'' by exerting a 
pressure on them and reducing their surface tension. Those nuclei are thus very different from those 
encountered on Earth. In order to make reliable predictions of the composition
of the inner crust, both nucleons bound inside ``clusters'' and free neutrons have to be described consistently. 
This can still be done using our effective force underlying our mass models by solving the 
HFB equations~(\ref{5}) with Bloch boundary conditions
\begin{eqnarray}
\label{9}
\psi^{(q)}_{1i}(\pmb{r}+\pmb{\ell}, \sigma)=\exp({\rm i} \pmb{k}\cdot\pmb{\ell})\psi^{(q)}_{1i}(\pmb{r}, \sigma)\nonumber\\
 \psi^{(q)}_{2i}(\pmb{r}+\pmb{\ell}, \sigma)=\exp({\rm i} \pmb{k}\cdot\pmb{\ell})\psi^{(q)}_{2i}(\pmb{r}, \sigma) 
\end{eqnarray}
where $\pmb{k}$ is the Bloch wave vector and $\pmb{\ell}$ is any lattice vector. 
But such calculations are computationally very expensive. So far self-consistent mean field calculations have
been performed using a simpler approach based on the Wigner-Seitz (W-S) method with~\cite{ne73} and without 
pairing~\cite{bal07}. However the W-S treatment introduces spurious neutron shell effects which contaminate 
the results~\cite{bal06,cha07,mar08}. 
For this reason, we have followed a different strategy by applying the Extended Thomas-Fermi method including 
proton shell corrections via the Strutinsky integral (see Ref.~\cite{onsi08} for details). This ETFSI method is not 
only a very fast approximation to Hartree-Fock equations, but it also avoids the pitfalls of boundary 
conditions than plagued current quantum calculations. This method could be similarly generalized 
to solve approximately the HFB equations~(\ref{5}). But in our calculation of the equation of state~\cite{onsi08}, 
pairing was neglected. Pairing is not 
expected to have any significant impact on the energy density and on the pressure because it only affects nucleon 
states lying close to the Fermi level. Indeed in uniform neutron matter, using our model HFB-16~\cite{cha08}
we have found that the pairing contribution represents at most $\simeq0.5$\% of the energy per particle (without rest mass energy).
But of course pairing is essential for studying neutron superfluidity. In the bottom layers of the crust where spatial inhomogeneities are small, the effects of the nuclear clusters on the neutron superfluid can be estimated by solving 
the HFB equations perturbatively. Any field $\phi(\pmb{r})$ having the periodicity of the crystal lattice 
(i.e. single-particle fields, pairing field) can be expanded into Fourier series
\begin{equation}
\label{10}
\phi(\pmb{r})=\widetilde{\phi}_0 + \sum_{\pmb{G}\neq0} \widetilde{\phi}_{\pmb{G}}\exp({\rm i}\, \pmb{G}\cdot\pmb{r})
\end{equation}
where $\pmb{G}$ are reciprocal lattice vectors. The Fourier coefficients are defined by 
\begin{equation}
\label{11}
\widetilde{\phi}_{\pmb{G}}=\frac{1}{V_{\rm cell}}\int_{\rm cell} {\rm d}^3\pmb{r}\, \phi(\pmb{r})\exp(-{\rm i}\, \pmb{G}\cdot\pmb{r})
\end{equation}
with $V_{\rm cell}$ the volume of the unit cell. If $\phi(\pmb{r})$ is spatially slowly varying, we will have 
$|\widetilde{\phi}_{\pmb{G}}|\ll |\widetilde{\phi}_0|$ for any $\pmb{G}\neq0$.
Solving to lowest order the HFB equations~(\ref{5}) for neutrons with Bloch boundary conditions~(\ref{9}) 
thus leads to the gap equation
\begin{equation}
\label{12}
\widetilde{\Delta}_n=- \frac{1}{8\pi^2} \widetilde{B}_n^{3/2} \widetilde{v^{\pi\, n}} \, \widetilde{\Delta}_n \,\int_\Lambda {\rm d}\varepsilon 
\frac{\sqrt{\varepsilon}}{\sqrt{(\varepsilon-\widetilde{\lambda}_n)^2+\widetilde{\Delta}_n^2}} \, ,
\end{equation}
where $\widetilde{B}_n$ and $\widetilde{v^{\pi\, n}}$ are given by
\begin{equation}
\label{13}
\widetilde{B}_n=\frac{1}{V_{\rm cell}}\int_{\rm cell} {\rm d}^3\pmb{r}\, B_n(\pmb{r})\, , \hskip 0.5cm\widetilde{v^{\pi\, n}}=\frac{1}{V_{\rm cell}}\int_{\rm cell} {\rm d}^3\pmb{r}\, v^{\pi\, q}[\rho_n(\pmb{r}),\rho_p(\pmb{r})]\, .
\end{equation}
This equation is similar to the BCS Eq.~(\ref{8}) after substituting $B_n$ and $v^{\pi\, n}$ by their spatial average. This result
is an illustration of the proximity effect: all particles whether inside clusters or not contribute to the pairing gap.
Since the pairing gap is typically very small compared to the Fermi energy, we can approximate the chemical potential by
the latter
\begin{equation}
\label{14}
\lambda_n=\widetilde{B}_n k_{{\rm F}n}^2\, , \hskip0.5cm k_{{\rm F}n}=(3\pi^2 \rho_n)^{1/3}\, .
\end{equation}
We have calculated the neutron pairing gap $\widetilde{\Delta}_n$ in the densest layers of the inner crust of neutron stars 
using the nucleon density profiles obtained with the ETFSI method. In order to study the modifications of the pairing gap 
due solely to the presence of spatial inhomogeneities, we have applied our model HFB-16~\cite{cha08} which was adjusted 
on the $^1S_0$ pairing gap of pure neutron matter, as calculated with realistic forces but without any medium effects. 
For comparison, we have also calculated the pairing gap $\Delta_n$ of uniform neutron matter for the density $\rho_n^f$ corresponding 
to the density of free neutrons. Results are summarized in Table~\ref{tab1}. We have found that nuclear clusters reduce 
the neutron pairing gap $\widetilde{\Delta}_n$ compared to $\Delta_n$. This can be understood by the fact that the neutron 
pairing gap arises from the spatial average of the pairing strength which is smaller inside clusters than outside. 

\begin{table}[t]
\caption{Equilibrium composition of the bottom layers of neutron-star crust and $^1S_0$ neutron pairing gaps. 
$\rho$ is the average nucleon density, $Z$ and $A$ the equilibrium numbers of protons and 
nucleons in the W-S cell respectively (as obtained from the ETFSI method~\cite{onsi08} with effective force BSk16~\cite{cha08}), 
$\rho_n^f$ the neutron density outside clusters, $\Delta_n$ is the $^1S_0$ pairing gap 
of uniform neutron matter for the neutron density $\rho_n^f$, $\widetilde{\Delta_n}$ is the neutron pairing gap
obtained by solving the HFB equation perturbatively.}
\setlength{\tabcolsep}{9.7pt}
\label{tab1}
\begin{tabular*}{\textwidth}{@{\extracolsep{\fill}}c|c|c|c|c|c}
$\rho$ [fm$^{-3}$] & $Z$ & $A$ & $\rho_n^f$ [fm$^{-3}$] & $\Delta_n$ [MeV] & $\widetilde{\Delta_n}$ [MeV]\\
\hline
0.070 & 40 & 1258 & 0.060 & 1.79 & 1.48 \\ 
0.065 & 40 & 1264 & 0.056 & 1.99 & 1.72 \\ 
0.060 & 40 & 1260 & 0.051 & 2.20 & 1.96 \\ 
0.055 & 40 & 1294 & 0.047 & 2.40 & 2.21 \\ 
0.050 & 40 & 1304 & 0.043 & 2.59 & 2.45 \\ 
\end{tabular*}
\end{table}

\section[]{Conclusions}
                 
Our HFB-17 mass model~\cite{gcp09} not only gives a better fit to the mass data than any 
other mean-field model, but is also by far the most microscopically founded.
With the additional constraints on the properties of pure neutron matter, this model is 
thus expected to make more reliable predictions of highly neutron-rich nuclei. Besides it 
is very well suited for the description of astrophysical environments like supernova cores 
and neutron stars. In particular, this is the first of our models well adapted to 
the investigation of a possible superfluid phase in the inner crust of neutron stars. 

\section*{Acknowledgments}
The financial support of the FNRS (Belgium) and the NSERC 
(Canada) is acknowledged.

\end{document}